\documentclass[twocolumn,superscriptaddress]{revtex4-1}
\usepackage{color}
\usepackage{graphicx}
\usepackage{amssymb}
\usepackage{amsmath}
\usepackage{epstopdf}
\usepackage{natbib}
\usepackage{hyperref}

\begin{document}

\title{A tractable genotype-phenotype map for the self-assembly of protein quaternary structure}

\author{Sam F. Greenbury}
\affiliation{Theory of Condensed Matter Group, Cavendish Laboratory, University of Cambridge, UK}
\author{Iain G. Johnston}
\affiliation{Department of Mathematics, Imperial College London, London, UK}
\author{Ard A. Louis}
\affiliation{Rudolf Peierls Centre for Theoretical Physics, University of Oxford, UK}
\author{Sebastian E. Ahnert}
\affiliation{Theory of Condensed Matter Group, Cavendish Laboratory, University of Cambridge, UK}

\date{\today}
\begin{abstract}
The mapping between biological genotypes and phenotypes is central to the study of biological evolution. Here we introduce a rich, intuitive, and biologically realistic genotype-phenotype (GP) map, that serves as a model of self-assembling biological structures, such as protein complexes, and remains computationally and analytically tractable. Our GP map arises naturally from the self-assembly of polyomino structures on a 2D lattice and exhibits a number of properties: \textit{redundancy} (genotypes vastly outnumber phenotypes), \textit{phenotype bias} (genotypic redundancy varies greatly between phenotypes), \textit{genotype component disconnectivity} (phenotypes consist of disconnected mutational networks) and \textit{shape space covering} (most phenotypes can be reached in a small number of mutations). We also show that the mutational robustness of phenotypes scales very roughly logarithmically with phenotype redundancy and is positively correlated with phenotypic evolvability. Although our GP map describes the assembly of disconnected objects, it shares many properties with other popular GP maps for connected units, such as models for RNA secondary structure or the HP lattice model for protein tertiary structure.
The remarkable fact that these important properties similarly emerge from such different models suggests the possibility that universal features underlie a much wider class of biologically realistic GP maps.
\\
\end{abstract}

\maketitle


\section{Introduction}
Evolution is one of the most fundamental principles in biology. While organismal genotypes are becoming accessible due to rapid advances in sequencing technology, further understanding of the complicated mapping from sequence to phenotype is necessary for a richer understanding of evolutionary dynamics \citep{Alberch_1991, Wagner_book, Noble_book, Pig_2010}.
While the terms genotype and phenotype can be flexibly assigned in a biological system, genotypes are generally defined as the genetic material upon which mutations act and phenotypes capture the properties of the organism on which selection can differentiate. As such, the mapping from genotype to phenotype -- the \emph{GP map} -- links mutations to potentially selectable variation, and is therefore of critical importance in understanding evolutionary systems. GP maps also provide a basis for understanding important biological concepts such as mutational robustness and evolvability, which may profoundly affect evolutionary dynamics, and help determine the fundamental topologies of the landscapes upon which evolutionary processes occur \citep{Svensson_2012}.

In general it is intractable to directly model the details of even small parts of a whole organismal GP map, due to both the very large numbers of genotypes, and a lack of knowledge of all possible phenotypes. Advances have been made in recent years, however, with the use of simplified models. Three particular systems have been modelled with notable success. Firstly, genetic regulatory networks have been approximated using a variety of abstract models, including Boolean networks \citep{Kauffman_1969,Li_2004}. Despite their simplicity, Boolean networks have demonstrated a remarkable ability to produce biologically realistic results. For example, they have been shown to reproduce key aspects of the yeast cell cycle \citep{Li_2004}. Secondly, RNA secondary structure can be routinely and accurately predicted via a host of different methods \citep{Mathews_2010}, as a result of which it has become one of the best-known GP maps \citep{Schuster_1994,Pig_2010}, particularly for the study of evolutionary dynamics \citep{Schuster_1994, Cowperthwaite_2007}. Finally, the complex problem of a protein folding into well defined tertiary structure has been investigated using various models including the highly simplified HP lattice model, where folds are represented as self-avoiding walks on a lattice, and the full sequence is reduced to binary alphabet (H stands for hydrophobic and P for polar amino acids) \citep{Dill_1985}. Despite its heavily coarse-grained nature, this model has produced important biological insights \citep{Li_1996} and has been shown to accurately model known features of protein tertiary structure \citep{BB_1997}. While these models have been successful for specific biological examples, another very important advantage of their tractability is the potential for extracting more general underlying principles of GP maps, building on our understanding of how evolution {shapes} the natural world \citep{Wagner_book, Pig_2010}.

\begin{figure*}
	\centering
	\includegraphics{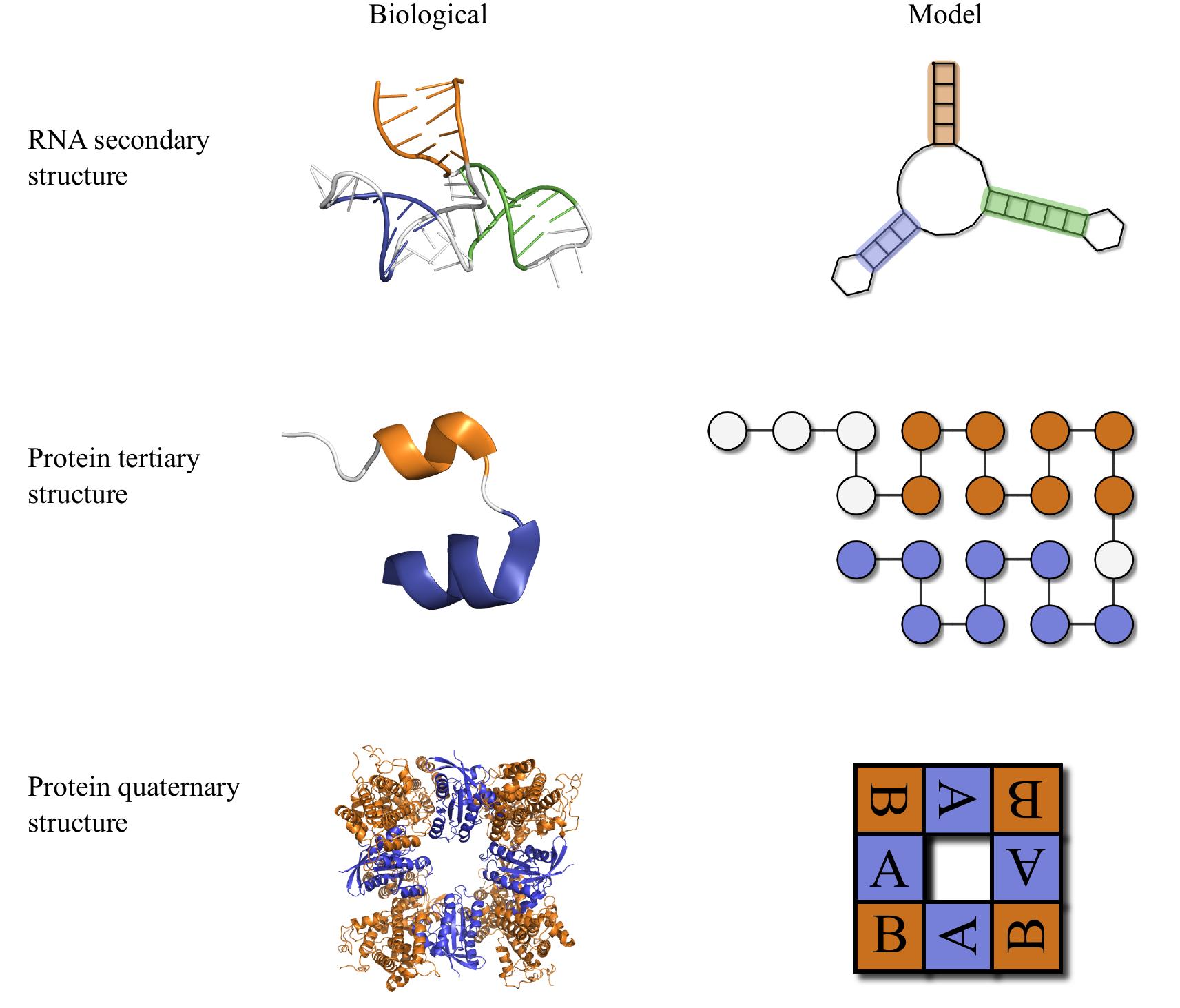}	
	\caption{A comparison of how  three different biological structures may be represented in a corresponding model system.  The top row compares a version of the iconic hammerhead ribozyme (PDB reference 1RMN \citep{1rmn_cite}) with its secondary structure  representation (showing the bonding pattern of nucleotides) produced by the Vienna RNA package \citep{Hofacker_2003} shown alongside.  The orange, blue and green colours in each part represent the bonded stems in the structure. The middle row depicts a cartoon of the tertiary structure of a single chain of length 21 (chain A) from an insulin protein (PDB reference 1APH \citep{1aph_cite}) which is compared to a schematic HP lattice protein interpretation on the right. The orange and blue colours are used to demonstrate the structural feature of alpha helices in the pictures. Finally in the bottom row, we show a protein complex (PDB reference 1BKD \citep{1bkd_cite}) alongside a polyomino representation. The orange and blue colours represent the different subunits involved in the protein. The ability of the polyomino representation to capture the C$_4$ symmetry of 1BKD is apparent from the rotation of the labelling on the subunits.}
	\label{comparisons}
\end{figure*}

In this paper we extend this family of coarse-grained models for biological structure, exploring a very recently introduced model for tile self-assembly \citep{Poly_2010,Poly_2011} which can be viewed as a highly coarse-grained  GP map for protein quaternary structure (protein complexes). Understanding the formation of protein complexes is important as demonstrated by the fact that most proteins form complexes in the cell (around $80\%$ in yeast \citep{Levy_2006}) and the function of these complexes is often strongly linked to their physical form. Protein complexes are formed by the interaction of multiple individual protein chains to form larger structures. The interaction between two chains is predominantly mediated by hydrophobic forces acting to pack together non-polar amino acids to provide fewer energetically unfavourable interactions with water \citep{Perica_2012}. Invaluable resources for the study of protein complexes are the Protein Data Bank, providing a database containing experimentally known protein quaternary structures \citep{pdb}, and the 3D complex database \citep{3d_complex} categorising PDB structures with a graph representation of the interactions between different subunits and a characterisation of the symmetry in a complex.
 
The relationship between the topology of a protein complex and the individual amino acids that make up its protein chains is highly complex due to the multiple functionality encoded in the protein sequence. For example, correct tertiary structure, folding pathways and other inter-protein interactions are all potential requirements for a single protein chain. Given these complexities, a direct and complete GP map of protein quaternary structure is intractable as including all required functionality would be unfeasible. Instead, in the spirit of the highly simplified HP model for protein tertiary structure, we represent the proteins as square tiles on a lattice \citep{Poly_2010,Poly_2011}.  Interactions between tiles model the protein-protein interactions that lead to self-assembly. In the model, a genotype is a sequence of characters describing interactions on the edges of the tiles, which, when combined with a self-assembly process on a 2D square lattice, leads to the formation of phenotypes comprising of different square tile building blocks conjoined along interacting edges. These square tile structures are known as polyominoes and thus, we refer to the model as the Polyomino model and the resulting GP map as the Polyomino GP map.  They are closely related to a wider class of  lattice tiling models that have a long and important history in mathematics and computer science (which we discuss further in Section \ref{section:model}).

Despite these great simplifications, and in analogy with the highly schematic HP model, RNA secondary structure models or Boolean network models of GRNs, we expect the Polyomino model to provide insight into the general structure of the full GP map for the formation of protein complexes from folded proteins. In fact, our earlier work \citep{Poly_2011} has already discussed the evolutionary dynamics of the Polyomino model, demonstrating that it can be used to rationalise the preference of dihedral over cyclic symmetries in homomeric protein tetramers \citep{Levy_2008, Villar_2009}.

In Fig. \ref{comparisons}, we compare coarse-grained representations for RNA secondary structure, protein tertiary structure and protein complexes. On the left hand side, the biological representation is shown and on the right, a given model representation. Through this figure we wish to highlight two points. Firstly, that each of the model systems is a dramatically simplified version of the corresponding biological system and that the Polyomino model is of a similar order of coarse-graining to previous models. And secondly, that the Polyomino model can provide a concise representation of real protein complexes by capturing features such as the symmetry of the subunit arrangements (C$_4$ in this case).
 
In contrast to RNA secondary structure or protein tertiary structure, where structure forms through the folding of a connected string of individual entities (nucleotides or amino acids),  protein complexes are built by joining separate individual entities (protein chains). Furthermore, while for string-like self-assembly the final structure size is constrained, proteins can form unbounded structures that can be ordered or disordered. In our work here, we focus on the subset of deterministic, finite sized structures to model protein quaternary structure.  But, in principle, the Polyomino model can also capture the more general phenomenology of unbounded and non-deterministic assembly \citep{Poly_2010,Poly_2011}. Moreover, tiling models have been used to study synthetic self-assembling systems \citep{DNA_sa}, and a better understanding of the design space of polyominoes may aid in the design of these artificial systems. 

This article proceeds as follows. We first describe in detail the Polyomino model and some of its fundamental properties (Section \ref{section:model}). We then analyse a wide range of properties of the resultant GP map and compare these properties to those described, for example in \citep{FW_2012}, for the HP model and the RNA secondary structure map.  In particular, we show in Sections \ref{section:redundancy}-\ref{section:components} that the mapping from sequences to phenotypes for all three models share  the following general properties:  \textit{redundancy} (there are many more  genotypes than phenotypes) leading to large \textit{neutral sets} (the collection of all genotypes that map to a given phenotype) and  \textit{phenotype bias} (some phenotypes are associated with many more genotypes than others).  A more fine-grained analysis shows that the neutral sets also exhibit \textit{component disconnectivity} (not all genomes in a neutral set  can be linked with single mutational steps). We proceed with a more detailed comparison of the Polyomino and RNA systems, through considering \textit{shape space covering} (most phenotypes can be reached from any other phenotype with just a small number of mutations), before showing the mean mutational robustness of a phenotype (the \textit{phenotypic robustness}) scales very roughly logarithmically with the redundancy of a phenotype, and finally that it is positively correlated with the \textit{evolvability} (defined here as the number of other phenotypes \textit{potentially} accessible from a phenotype), as postulated to hold more generally by \citep{Wag_re_paradox}. Finally, in Section \ref{section:discussion} we discuss some implications of the remarkable agreement we find between the structure of our Polyomino GP map and those of the better studied RNA secondary structure and HP maps.

\section{The polyomino self-assembly model and its associated GP map}\label{section:model}
\begin{figure*}
	\centering
	\includegraphics{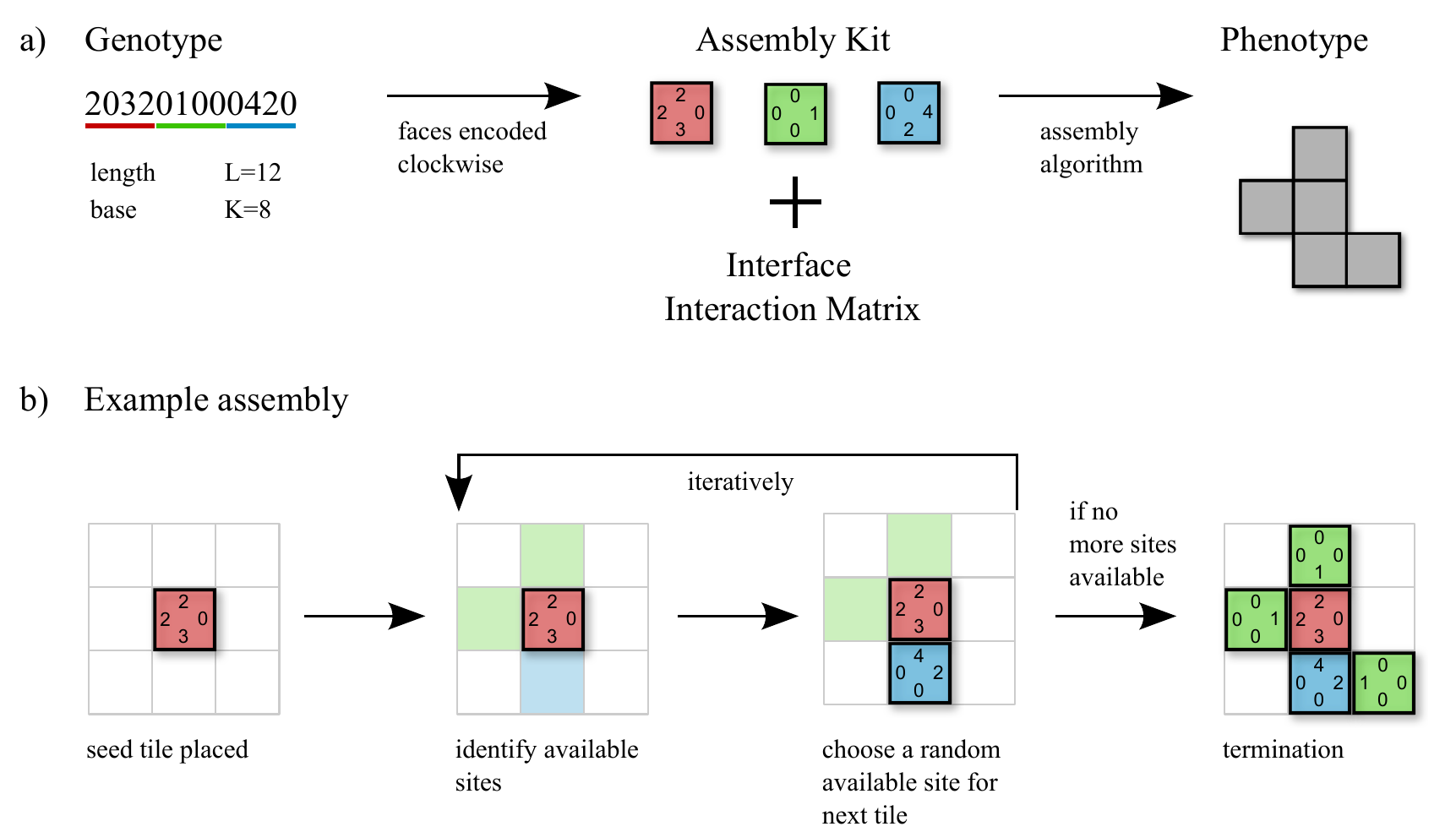}	
	\caption{The \textit{Polyomino GP map}. (a): The translation from an example genotype to phenotype is depicted, with the intermediate conversion into an assembly kit shown. In the genotype, the characters are underlined with the colour of the block they are assigned to. The interface interaction matrix is not shown explicitly, but throughout our work we assume the convention of interface types interacting in pairs ($1\leftrightarrow2$, $3\leftrightarrow4$...), with 0 and $c-1$ being neutral. The interaction matrix together with the assembly kit are passed to the assembly algorithm, which is used to produce the phenotype. (b): An example of the assembly process is shown, which proceeds in discrete time steps. The first tile in the assembly kit is used to \textit{seed} the assembly. Further time steps follow, with the identification of \textit{available sites} (depicted in light green and blue in the second picture from left), followed by the random choice of an available site and placement of the corresponding tile (a blue tile is placed on the blue available site in this example). Assembly is terminated when there are either no more available sites (as above), or if the structure is larger than $D_{max}$ (number of blocks) in width or height after which we assume the structure will no longer terminate but grow indefinitely.}
	\label{method}
\end{figure*}

The process of tiling and its connection with computer science was first developed by \citet{wang_tiles}. Since then, tiling models have been shown to be capable of computation and, in particular, Turing-universal computation under the condition that cooperative binding is allowed between tiles, demonstrating the ability of 2D tiling systems to model computational as well as structure-forming processes \citep{Winfree_1999}. \citet{Rothemund_2000} studied the program size complexity necessary to build a structure of a given size. More general considerations of the complexity of tiled structures have since been discussed in \citep{Solo_2007}, with a more biological slant given by \citet{Poly_2010} and applications to artificial biological systems discussed by \citet{DNA_sa}. 

Here, rather than focussing on these tiles as potential computing devices or as models for complexity, we explore how they can be used to understand the GP map of a specific biological system, namely the self-assembly of finite-sized protein structures.  Nevertheless, we are aware that some of our conclusions may have applications for a wider class of systems.

We now proceed with a more detailed description of the Polyomino model as a GP map.

\subsection{Summary of the Polyomino GP map}
The genotype is modelled as a character string representation of a set of $N_t$ tiles which make up an \textit{assembly kit}. The edges of each tile in the assembly kit is given a number which represents the \textit{interface type}. Interactions between interface types are defined via an \textit{interface interaction matrix} $A_{ij}$. In our work here, we only consider one such interaction matrix type, with a total of $c$ interface types. There are no self-interacting interface types and interface types interact in unique pairs ($1\leftrightarrow2$, $3\leftrightarrow4$, ...), with the only fully neutral types being $0$ and $c-1$. Defining interface interactions in this way allows the single parameter $c$ to control the number of potential unique bond types. As such, the two parameters that describe a given parameterisation of the Polyomino GP map are the tile number $N_t$ and the total number of interface types $c$, allowing a particular Polyomino GP map to be labelled as $S_{N_t, c}$.

Below we discuss the genotype, phenotype and map used in the Polyomino GP map.

\subsection{Genotype}
The genotype for the set of building blocks in the assembly kit is written as a bit-string of length $L$ in base $K$. The base we employ in this paper is the total number of interface types $c$ used in a given GP map. This allows each base to mutate to any other base at each site. The procedure of converting a genotype string into the assembly kit is part of the \textit{map}.

\subsection{Phenotype}
In the context of the self-assembly mapping, there are several ways of classifying a polyomino structure. These may include criteria based on its overall shape and the individual tile types making up the final polyomino structure, as well as individual tile orientations. These different possibilities are discussed in \citep{Poly_2011}.

Here, we will classify the phenotype according to the overall shape of the polyomino independent of origin translation and C$_4$ (90 degree) rotations. Note that chiral counterparts of polyominoes represent distinct phenotypes.

\subsection{Map}
The map has two stages: conversion of the genotype into the assembly kit, followed by assembly of a polyomino from an assembly process involving the assembly kit and the interface interaction matrix. A diagram representing this process is shown in Fig. \ref{method}.

\subsubsection{Genotype to Assembly Kit}
The characters of the genotype are read from left to right along the string. Characters are assigned to the next blank edge of a square tile in the assembly kit. The edges are taken in clockwise order (from the top side) with all edges being written before moving on to a new block. The total number of tiles, in terms of the genotype string length, can be expressed as $N_t = L/4$.

\subsubsection{Assembly Kit to Phenotype}
The assembly of the 2D polyomino takes place on a square lattice where individual tiles from the assembly kit are placed. The interface types on the edges of tiles can form an attractive interaction as determined by the binary interface interaction matrix $A_{ij}$, with 1 denoting attraction and 0 neutrality. A bond may form between tiles if two adjacent (interacting) edges have interface types which attract, as defined by the interaction matrix.

The assembly process is initialised by placing (seeding) a single tile on the lattice. We will consider only GP maps in which the seed tile corresponds to the first tile described in the genotype. A different protocol where any tile may be used to seed the assembly is also possible and does not significantly effect the results presented here. The assembly then proceeds as follows:
\begin{enumerate}
	\item{\textit{Available sites} on the lattice are identified. These are places on the lattice where a tile may be placed in a particular orientation such that it will form a bond to an adjacent tile that has already been placed. In the assembly algorithm, a list is kept of the position, tile type and orientation for each possible tile placement.}
 	\item{A random available site on the lattice is chosen.}
	\item{The chosen site is filled with the associated tile and with the corresponding orientation.}
\end{enumerate}

These steps are repeated until either:
\begin{itemize}
	\item{There are no available sites for bonding.}
	\item{The structure grows beyond a certain width or height $D_{\text{max}}$, which is taken as a proxy for unbounded assembly, so that the resulting phenotype is described as \textit{unbound}. We set $D_{\text{max}}=16$ here, but our results are not sensitive to this cutoff as, for the polyomino systems we study here, there are no bounded structures larger than this.}
\end{itemize}

At this point the assembly process is terminated and the structure produced is recorded. To test whether the structure is \textit{deterministic}, the assembly process is repeated $k$ times, with each C$_4$ rotation of the final structure checked against the recorded structure. If there are any differences between any of the $k$ assemblies, the phenotype is classified as \textit{non-deterministic}. Phenotypes that are classified as unbound or non-deterministic structures are represented by a single phenotype, which we refer to as the undetermined (\textit{UND}) phenotype and is assumed not to be biologically relevant in this context. A more detailed discussion of the classification of polyomino structures is given in \citep{Poly_2011}.

\section{Properties of the polyomino GP map}\label{section:gp_props}
In this section, we analyse the Polyomino GP map by making measurements of redundancy, phenotype bias and component numbers, before moving on to analyse the properties of shape space covering, robustness and the relationship between robustness and evolvability. For each measurement, comparisons are made with the RNA secondary structure model and, for the first three of these properties, an HP lattice protein model. The software used to model RNA secondary structure was the well-known Vienna package \citep{Hofacker_2003}, and for the HP lattice model we used the data from the spaces enumerated by \citet{Irback_2002}.

\subsection{Redundancy}\label{section:redundancy}
\begin {table}[t]
\centering
	\begin{tabular}{ l l r r r }
	\hline
	GP Map & $N_G$ & $N_P$ & $N_{cov}/N_P$ & $N_C$\\
	\hline
	Polyomino $S_{2,8}$ & $1.7\times10^7$ & 13 & 54\% & 1,347 \\
	Polyomino $S_{3,8}$ & $6.9\times10^{10}$ & 147 & 16\% & -- \\
	Polyomino $S_{4,16}$ & $1.8\times10^{19}$ & $^\ast$2,237 & 3\% & --\\
	RNA, $L=12$ & $1.7\times10^7$ & 57 & 47\%& 645\\
	RNA, $L=15$ & $1.1\times10^9$ & 431 & 23\% & 12,526\\
	RNA, $L=20$ & $1.1\times10^{12}$ & 112,118 &10\% & --\\
	HP, $L=25$ & $3.4\times10^7$ & 107,335 & 68\% & 148,253\\
	\hline
	\end{tabular}
\caption{\small{\textit{Redundancy}, \textit{phenotype bias} and \textit{components} in Polyomino, RNA and HP GP maps. Comparing the number of phenotypes ($N_P$) to the number of genotypes ($N_G$) for each GP map highlights large-scale redundancy present. Phenotype bias is demonstrated in each map with the measure of the fraction of phenotypes that covers 95\% of the genotypes ($N_{cov}$ is the number of phenotypes that covers the 95\% of genotypes). In all cases the fraction of phenotypes is significantly smaller than the fraction of genotypes being covered, indicating the presence of a strong phenotype bias. The final column is the total number of genotype components ($N_C$) in each GP map. In all cases (non-computable values left out) the number of components is larger than the number of phenotypes, indicating phenotypes tend to be spread out over multiple disconnected components. RNA data for $L=12$ was computed from the Vienna package \citep{Hofacker_2003} and taken from \citep{SS_phd} for $L=15,20$. The starred Polyomino $S_{4,16}$ value for $N_P$ is estimated from large-scale sampling of the GP map over multiple runs of the algorithm presented in \citep{Jorg_2008}. The HP results were calculated from the data made available by \citep{Irback_2002}. Non-deterministic phenotypes and the trivial structure in RNA are excluded from the statistics.}}
\label{redundancy}
\end{table}

It is now well established that many different sequences can generate similar protein or RNA phenotypes \citep{Wagner_book2}. This large-scale redundancy is the basis, for example, of the molecular clock hypothesis, which is widely used for inferring phylogenetic relationships \citep{Thorpe_1982}. It is therefore not surprising that  such redundancy (also known in the literature on GP maps as degeneracy/designability) has been observed in model RNA \citep{Schuster_1994}, HP lattice protein \citep{BB_1997} and genetic regulatory \citep{Ciliberti_2007} GP maps, as well as in more general model systems such as signalling networks \citep{Fernandez_2007}. As such, redundancy is expected to be a typical feature of GP maps.

In Table \ref{redundancy}, we show the number of genotypes ($N_G$) and phenotypes ($N_P$) for different polyomino, RNA and HP GP maps. As expected all three models show significant redundancy. The Polyomino model displays large-scale redundancy shown by the vastly fewer phenotypes in comparison to genotypes for each of the GP maps presented. As can be seen in Table \ref{redundancy}, this is of a similar order to the RNA GP maps. For example, for RNA $L=12$ and Polyomino $S_{2,8}$ (which both have $1.7\times10^7$ genotypes), there are $57$ phenotypes for the former and $13$ for the latter.

\subsection{Phenotype Bias}\label{section:bias}
\begin{figure}
\centering
	\includegraphics{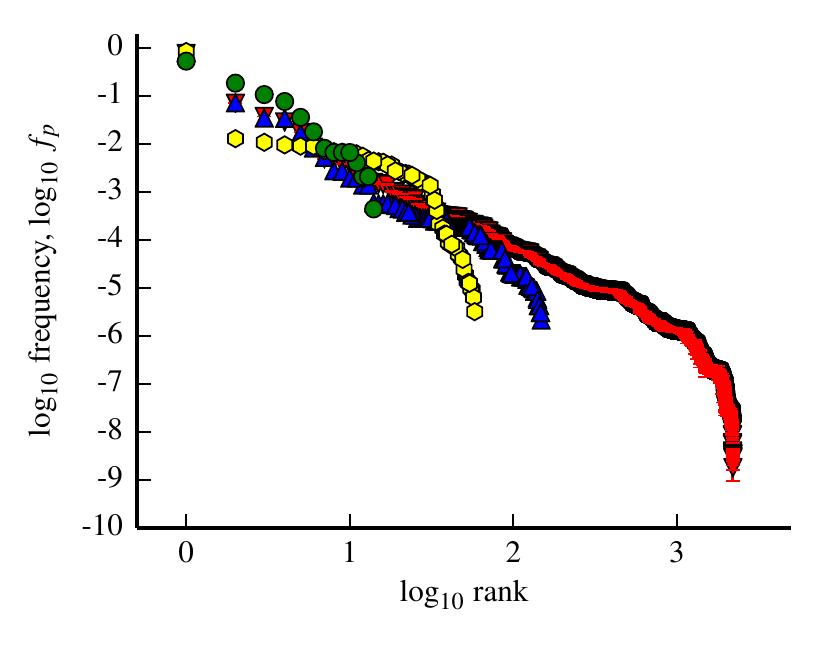}
	\caption{\small{\textit{Phenotype bias} in GP maps. The plot displays three Polyomino GP maps with increasing tile numbers $S_{2,8}$ (green), $S_{3,8}$ (blue) and $S_{4,8}$ (red) alongside the RNA $L=12$ GP map (yellow). Note the log-log scale. The frequency of a given phenotype (structure) is plotted against the rank of its frequency within the map. In all three Polyomino maps, we see an approximately exponential bias (linear trend in a log-log plot). The errorbars for the Polyomino $S_{4,16}$ map are the associated standard error on the mean for the estimates of frequencies calculated using the estimation algorithm for large maps developed by \citep{Jorg_2008}.}}
\label{bias}
\end{figure}

On top of the measure of redundancy in GP maps, it is also possible to consider \textit{phenotype bias}, the idea that some phenotypes are much more redundant than others. Examples of bias can again be seen in RNA \citep{Schuster_1994,Stitch_2008} and the HP model \citep{BB_1997,Li_1996,FW_2012}. We demonstrate bias in the GP maps in two ways. In Table \ref{redundancy} we present the fraction of phenotypes ($N_{cov}/N_P$, where $N_{cov}$ is the number of phenotypes) that cover 95\% of the genotype space. This statistic shows that in all three GP maps (RNA, HP and Polyomino), of the deterministic/non-trivial space, a smaller fraction of phenotypes is required to cover a given amount of the genotype space, indicating a bias in the number of genotypes assigned to each phenotype. In Fig. \ref{bias}, we plot the frequency (fractional coverage of genotype space) of each phenotype against its frequency-rank in three different Polyomino GP maps, as well as for the RNA $L=12$ GP map. In each of the Polyomino GP maps we see a large, approximately exponential bias in phenotype frequencies (linear trend in a log-log plot). The data from the RNA GP map also shows a distinct bias, although the relationship is more complicated than the more apparent exponential form seen in the Polyomino GP maps. Taken together these statistics show a significant amount of phenotype bias in the Polyomino GP map along with the RNA and HP GP maps.

\subsection{Component disconnectivity}\label{section:components}
Further to the consideration of redundancy and bias of phenotypes in genotype space, we can ask how many components a given phenotype forms in a particular GP map. In graph theory, a component is a set of nodes that are connected, and thus when applied to genotype spaces, is a set of connected genotypes with the same phenotype. Here we consider only point mutations. In RNA simple biophysical considerations show that two point mutations are needed to turn a purine-pyramidine  bond into a pyramidine-purine bond. This neutral reciprocal sign epistasis  leads to many individual components ~\citep{Aguirre_2011,SS_2012}.
 We expect similar behaviour in polyominoes, where there may be several ways of encoding a given phenotype that are not connected through neutral mutations due to there being multiple interface types.

In the final column of Table \ref{redundancy}, we show the number of genotype components ($N_C$) for four different GP maps: Polyomino $S_{2,8}$, RNA $L=12$, RNA $L=15$ and HP $L=25$ (Polyomino $S_{3,8}$ and $S_{4,16}$ and RNA $L=20$ GP maps were unobtainable due to computational expense). In all four we see a substantially larger number of components than phenotypes, indicating that phenotypes tend to form disconnected components. Although many components may be small, it is still the case  that whole-phenotype properties need to be considered carefully in the context of evolutionary dynamics, as it is not typically possible to access every genotype of a given phenotype through neutral mutation.

\subsection{Shape space covering}\label{section:ss_covering}
GP maps typically exist in high-dimensional genotype (sequence) spaces, a property with counterintuitive consequences and a topic of current biological research for GP maps \citep{Kauffman_1987, Gravner_2007}.  For example, for an alphabet of $K$ letters (e.g.\ $K=4$ for RNA or $K=20$ for proteins) the number of genotypes scales exponentially as $K^L$, while the number of mutations needed to reach any two genotypes is at most $L$, and so scales linearly. In practice, considerably fewer than $L$ mutations are needed to connect any two phenotypes.  This property has been studied most extensively for RNA GP maps, and in that context \citet{Schuster_1994}, borrowed the term shape space covering from its original use in immunology \citep{Perelson_1979}. The HP model has also been considered with respect to shape space covering \citep{BB_1997,FW_2012} with seemingly less rapid space covering.

We explore shape space covering in a similar way to \citet{FW_2012} who compared RNA and HP models, building on work of earlier investigators. Shape space covering is measured by counting the average fraction of phenotypes found when applying a given number of independent mutations (the radius, $n$) to a genotype in the GP map. This process involves picking a sample of genotypes and then measuring the number of phenotypes found in the ball of genotypes $n$-mutations around each of them. \citet{FW_2012} found that in both the RNA secondary structure and HP lattice protein GP maps, the fraction of phenotypes discovered increases in a roughly sigmoidal fashion with the increasing number of independent mutations applied to the source genotype and at a greater rate for RNA than the HP GP map.

\begin{figure}[t]
	\centering
	\includegraphics{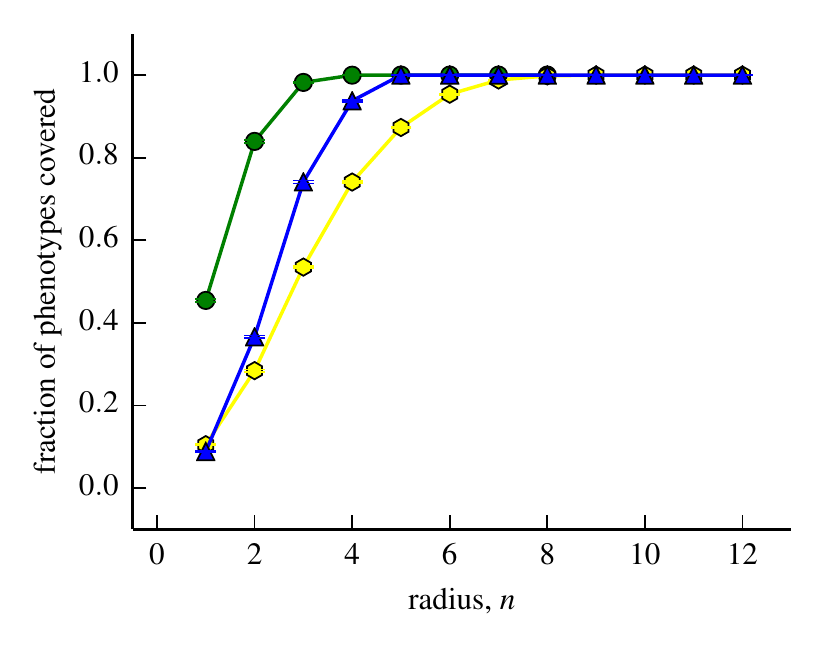}
	\caption{\small{\textit{Shape space covering} in Polyomino $S_{2,8}$ (green),  $S_{3,8}$ (blue) and RNA $L=12$ (yellow) GP maps. A random sample of 100 genotypes for each phenotype are taken, and the fraction of phenotypes discovered in a ball of radius $n$ is measured. This is then averaged over all sample phenotypes to give an approximate phenotype space covering for a given GP map. Both Polyomino and RNA GP maps exhibit similar behaviour -- shape space is almost completely covered in only a few mutations highlighting that in highly redundant spaces, most phenotypes can be reached in only a few mutations. Error bars are the standard error on the mean. The expected asymptotic value of unity is used for the Polyomino $S_{3,8}$ GP map for $n>4$ due to computational unfeasibility.}}\label{shape_space}
\end{figure}

In our work we measure shape space covering in a similar manner: we picked a sample of $1\text{,}000$ genotypes uniformly across all phenotypes and measured the phenotype of genotypes at each given radius ($n$). In Fig. \ref{shape_space} we plot the average fraction of phenotypes found for the Polyomino $S_{2,8}$, $S_{3,8}$ and RNA $L=12$ GP maps. The general behaviour of both the Polyomino and RNA GP maps is similar. Phenotypes are almost completely covered within half the sequence length, and at a slightly higher rate in the Polyomino GP maps. This provides clear evidence to suggest that the Polyomino GP map has shape space covering -- most phenotypes can be found within a ball containing many fewer genotypes than the entire space.

\subsection{Robustness}\label{section:robustness}
Biological systems need to be robust and consistently produce the same phenotype in response to environmental perturbations or to genetic mutations \citep{Wagner_book}. Here we consider only the phenotypic effects of mutations to the genotype. Mutational robustness describes the invariance of a phenotype as a result of mutations to the genotype. We focus here on single point mutations such that the fraction of 1-mutants that have the same phenotype as the original genotype under examination is defined as the \textit{genotypic robustness}. For a genotype $g$, with phenotype $p$, the genotype robustness can thus be expressed as follows

\begin{equation}\label{eq:r_g}
\rho_g = \frac{n_{p,g}}{(K-1)L}
\end{equation}
where $\rho_g$ is the genotypic robustness of $g$, $n_{p,g}$ is the number of 1-mutant neighbours of $g$ with phenotype $p$, $K$ is the base and $L$ is the sequence length (there are a total of $(K-1)L$ 1-mutants for any genotype). The robustness of the phenotype can then be considered as the average of this quantity over all genotypes with phenotype $p$, resulting in
\begin{equation}\label{eq:r_p}
\rho_p = \frac{1}{|\mathcal{P}|}\sum_{g\in\mathcal{P}}\rho_g
\end{equation}
where $\rho_p$ is the \textit{phenotypic robustness} and $\mathcal{P}$ is the set of genotypes with phenotype $p$. 

\begin{figure}[t]
	\centering
	\begin{tabular}{c}
	\includegraphics{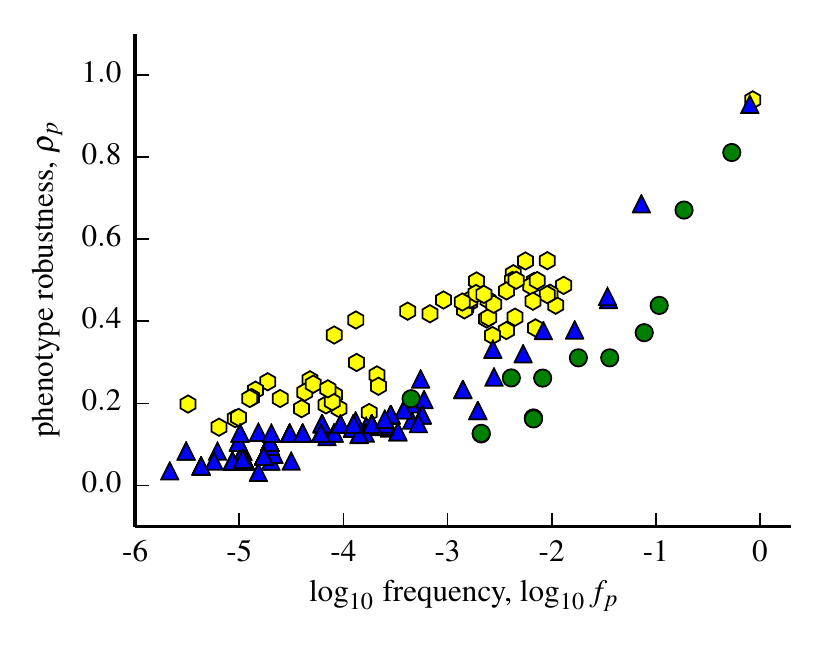}\\
	\end{tabular}
	\caption{\small{\textit{Robustness} in GP maps. Phenotypic robustness is plotted against frequency (log-scale) for the Polyomino $S_{2,8}$ (green), $S_{3,8}$ (blue) and RNA $L=12$ (yellow) GP maps. The top three points in the right hand side of the plot are the UND/trivial structure in the respective Polyomino and RNA systems. In the RNA case, the robustness scales roughly linearly with the logarithm of phenotype frequency. Whilst the trend is very roughly linear for the Polyomino GP maps, there is still a strong positive scaling relationship, demonstrating that polyomino phenotypes exhibit phenotypic robustness in a similar manner to the RNA GP map.}}\label{robustness}
\end{figure}

In Fig. \ref{robustness} we plot the phenotypic robustness $\rho_p$ against the frequency of each phenotype $f_p$ for the two Polyomino GP maps $S_{2,8}$ and $S_{3,8}$ and the RNA $L=12$ GP map.
Corroborating previous results \citep{SS_phd}, it can be seen that the RNA phenotypic robustness scales roughly logarithmically with phenotype frequency. This linear trend with log-frequency is very approximate for the Polyomino GP maps, although the robustness can still be seen to strongly scale with phenotype frequency. Both Polyomino GP maps exhibit a slightly smaller phenotypic robustness at a given frequency when compared to the RNA GP map. Nonetheless, despite these two observations, the clear indication here is that Polyomino GP maps have a phenotype robustness that scales similarly to phenotypes in the RNA GP map.

\subsection{Robustness versus Evolvability}\label{section:re}
\begin{figure*}
\centering
	\includegraphics{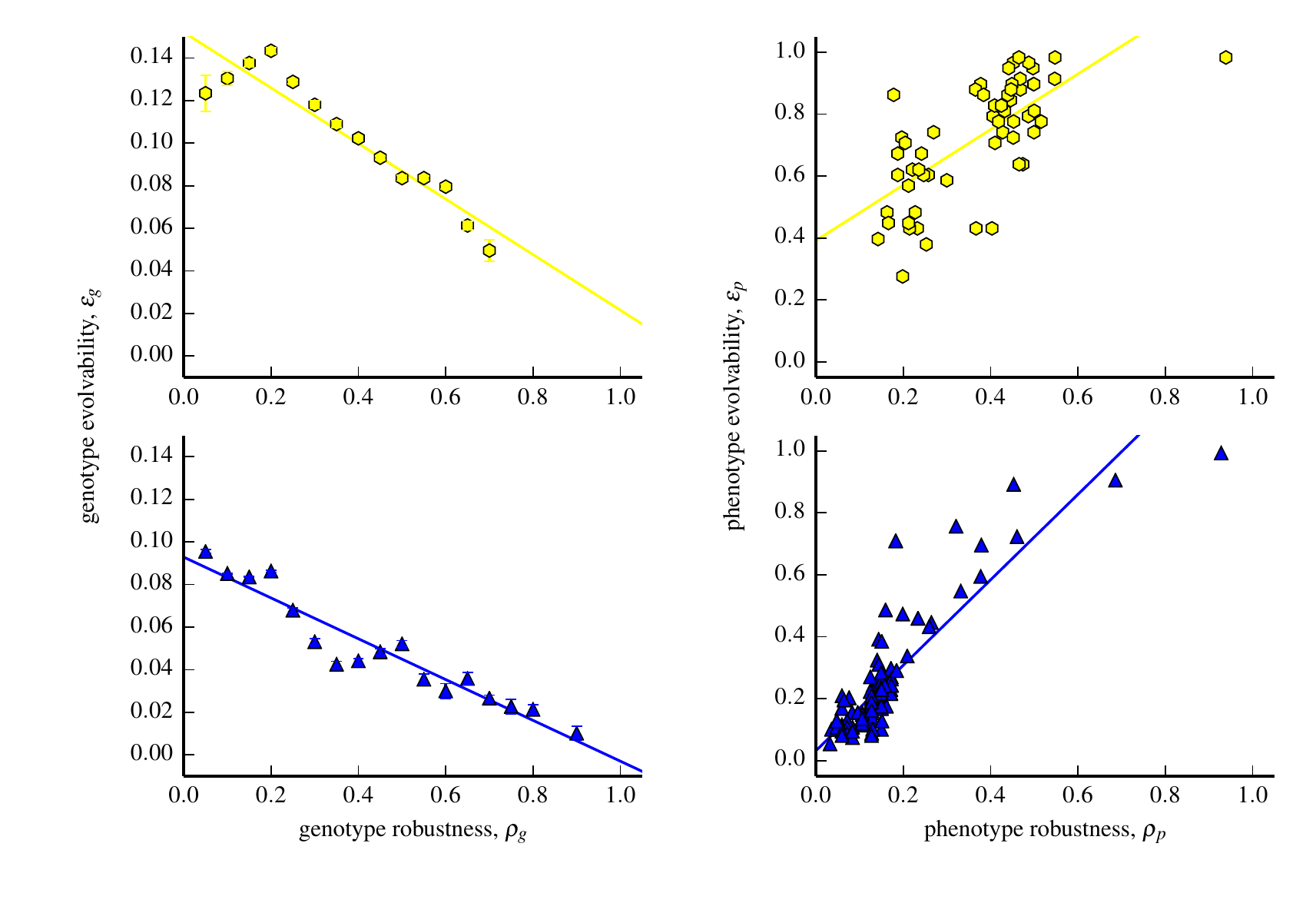}
	\caption{\small{\textit{Robustness} and \textit{Evolvability} in the RNA and Polyomino GP map. In the LEFT columns we show a significant negative correlation between genotypic robustness and genotypic evolvability, $\epsilon_g$ (RNA (yellow, top): p-value = 4.9e-07, $r^2$ = 0.89 and Polyomino (blue, bottom): p-value = 2.4e-09, $r^2$ = 0.91). The error bars are the standard error on the mean genotypic evolvability for each genotypic robustness bin. The more robust a genotype is, the fewer phenotypes lie in its 1-mutant neighbourhood. In the RIGHT column, there is a significant positive correlation between phenotypic robustness and phenotypic evolvability, $\epsilon_p$ (RNA: p-value = 1.5e-09, $r^2$ = 0.48 and Polyomino: p-value $<$ 2.2e-16, $r^2$ = 0.74 ). The far right hand phenotypes are the trivial structure and UND phenotype in RNA and Polyomino systems respectively. Both these results are in correspondence with \citep{Wag_re_paradox}.}}\label{re}
\end{figure*}

Robustness and evolvability are two evolutionary properties that have received much recent attention in the literature \citep{Kirschner_1998, Wag_re_paradox, Wagner_book, Bloom_2007, Gallagher_thesis, Masel_2010}. As discussed in the previous section, robustness is the ability of an organism to maintain its phenotype if its genotype is mutated. Evolvability on the other hand is considered as the capacity for producing phenotypic variation \citep{Alberch_1991, Wag_re_paradox}. One might expect that for an individual to be robust it would have to compromise its ability to produce variation (to be evolvable). Using the RNA GP map, \citet{Wag_re_paradox} demonstrated that this expected trade-off exists for individual genotypes, but that the two properties are in fact positively correlated at the level of phenotypes. A possible geometric explanation is that phenotypes can become more mutationally robust as a result of increased redundancy (more neutral neighbours). This in turn leads to connected components spreading out further across genotype space and  possessing a greater `surface', thus allowing a greater number of phenotypes to be accessible through neutral mutations. 

Wagner's argument is really about the static structure of genotype space and does not account for any dynamical evolutionary effects. Other authors have considered further static properties \citep{Cowperthwaite_2008}, including a different notion of phenotypic evolvability (diversity evolvability) defined as being the probability that two non-neutral mutations lead to different phenotypes. Such a definition of evolvability was found to not be correlated with robustness in the RNA GP map. A dynamical study  by \citet{Draghi_2010}  demonstrated that the relationship between robustness and evolvability could also depend on population parameters and mutation rates. In our work here, we simply consider whether the static properties of the Polyomino GP map behave in a similar manner to those of the RNA GP map as demonstrated by \citet{Wag_re_paradox} without here commenting on the wider debate of how robustness and evolvability relate.

In the left hand plots of Fig. \ref{re}, we show fractional evolvability versus fractional robustness, for genotypes and phenotypes in both the RNA $L=12$ and Polyomino $S_{3,8}$ GP maps. The genotype plot is a binned version of $100$ randomly sampled genotypes for each phenotype (apart from the non-deterministic/trivial structure) in each GP map. The fraction of all phenotypes that can be produced in any 1-mutation to each sampled genotype (genotypic evolvability, $\epsilon_g$) is plotted against the fraction of those 1-mutations that produce the same phenotype as that of the genotype being tested (genotypic robustness, Eq.\ref{eq:r_g}). For both polyominoes and RNA we see a significant negative correlation. This expected negative correlation simply represents the trade-off between genotypic evolvability and robustness at the level of the individual genotype

In the right hand plots of Fig. \ref{re}, we plot phenotypic evolvability against phenotypic robustness. The phenotypic evolvability  $\epsilon_p$ of phenotype $p$ is defined here as the fraction of all phenotypes that can be reached in a single mutation from any genotype with phenotype $p$. We also refer to this as the \textit{potential} evolvability because it represents the potential number of phenotypes that could be reached.  Whether they can be reached depends, of course, on details of the evolutionary dynamics.  For example, if only single mutations are available then $\epsilon_p$ should really be defined with respect to the relevant component~\citep{SS_2012}. Phenotypic robustness is defined in the same way as in Section \ref{section:robustness}, as the average genotypic robustness over all genotypes with phenotype $p$ (Eq. \ref{eq:r_p}). In the plots, we see strong positive correlations for both the RNA $L=12$ and the Polyomino $S_{3,8}$ GP maps, as expected. In other words, for both maps, phenotypes with a greater redundancy can be generated by a larger number of genotypes and are therefore, on average, more likely to be mutationally robust. Furthermore, such phenotypes are also likely to have a larger set of genotypes, and therefore a greater diversity of other phenotypes potentially accessible from this set. Thus, phenotypic robustness and potential evolvability are positively correlated.

\section{Discussion}\label{section:discussion}
In this paper, we have explored the properties of a GP map for biological self-assembly of the kind exhibited by protein quaternary structure based on a recently introduced Polyomino model for tile assembly~\citep{Poly_2010, Poly_2011}. We compared its properties to models of RNA secondary structure  and the HP model for protein tertiary structure. As is the case for these two well studied GP maps, we argue that even though our Polyomino model is highly schematic and thus misses many details of protein quaternary structure, it may nevertheless provide important biological insight  into the structure of the design space for protein complexes.

Despite the great complexity in potential phenotypes, the  polyomino model remains tractable as demonstrated in this paper by the ability to perform a wide variety of useful measurements on the GP map. Our main results are: 
 Firstly, the Polyomino model exhibits large-scale redundancy, a strong phenotype bias and the presence of disconnected genotype components across several parameterisations of the Polyomino GP map (Section \ref{section:redundancy}-\ref{section:components}). Secondly, that shape space may be covered in only a fraction of mutations -- that is, all phenotypes are a significantly smaller number of mutations away from each other than the total sequence length (Section \ref{section:ss_covering}). Thirdly, phenotypic robustness scales very roughly with the logarithm of the phenotype frequency (Section \ref{section:robustness}). And finally, genotypic robustness and genotypic evolvability are negatively correlated, whilst phenotypic robustness and phenotypic (potential) evolvability are positively correlated (Section \ref{section:re}). 
 
The Polyomino model describes the self-assembly of disconnected units (proteins) into finite sized structures (protein clusters) that can vary in size.  By contrast, for  RNA secondary structure and the HP model for protein tertiary structure, strings of connected units (nucleotides or amino acids) assemble into shapes of a fixed size.  Given the substantially different class of the phenotypes in our model, it is  remarkable that the measured properties of these GP maps turn out to be so similar. This begs the question of whether what we observe is in fact a more general property of self-assembling systems, or even broader, whether a wider class of GP maps will share these properties. This question can be sharpened by looking at the different properties separately. Redundancy should be widely shared across GP maps. Phenotype bias has been observed in models for gene-regulatory networks \citep{Raman_2011} and developmental networks~\citep{Borenstein_2008}. Could it be a more general property of GP maps? Disconnected components have also been observed in a Boolean threshold model for gene regulation \citep{Boldhaus_2010}. To our knowledge shape space covering has not been studied for other GP maps, but general considerations based on the high dimensionality of genetic space suggest that something like this may be more widely relevant \citep{Gavrilets_book}.  Finally, the correlation between phenotypic robustness and potential evolvability is deeply connected to the geometry of neutral sets and so is likely to be a much more general property of GP maps.  How this correlation plays out for realistic evolutionary dynamics is, of course, a much more complicated question.

 We are hopeful that more complete answers will be derived through further analysis and comparisons of different model GP maps in a similar manner to the work here and in \citep{FW_2012} or \citep{Wagner_book}. An important related question is whether model GP maps for parts of systems can be combined to achieve a more complete understanding of the evolution of phenotypic traits in the full organismal GP maps \citep{Wagner_book2}.

The Polyomino model can easily be adapted to study unbounded assembly or the assembly of synthetically produced objects like DNA tiles or patchy colloids \citep{DNA_sa}.  Thus the perspective gained from viewing polyominoes as a GP map may also  shed light on the artificial design process for these systems.

Finally, although we introduce the Polyomino model as a coarse-grained model for protein quaternary structure,  it is clear that the model is not capable of modelling particular intricacies of some individual proteins. For example, F1-ATPase is a protein whose subunit structure could not be accurately represented with a square tile model.  We argue that the Polyomino model nevertheless provides biological insight into questions about the global nature of the GP map that would not be computationally accessible using more complex models with greater biological detail. However, further work is needed to assess how well this new GP map performs in this respect.

\bibliographystyle{unsrtnat}

\end{document}